\documentclass[twocolumn,aps,prc,showpacs,superscriptaddress,preprintnumbers,floatfix,nofootinbib]{revtex4}
\usepackage{epsfig,graphics}
\usepackage{graphicx}
\usepackage{dcolumn}
\usepackage{bm}
\usepackage{amsmath}
\usepackage[usenames]{color}
\usepackage{ulem} 

\begin{document}

\title{Unified Description of Nuclear Stopping in Central Heavy-ion Collision from 10$A$ MeV to 1.2$A$ GeV}

\author{G. Q. Zhang}
\affiliation{Shanghai Institute of Applied Physics, Chinese
Academy of Sciences, Shanghai 201800, China} \affiliation{Graduate
School of the Chinese Academy of Sciences, Beijing 100080, China}
\author{Y. G. Ma \footnote{Corresponding author. E-mail address: ygma@sinap.ac.cn }}
\affiliation{Shanghai Institute of Applied Physics, Chinese
Academy of Sciences, Shanghai 201800, China}
\author{X. G. Cao}
\affiliation{Shanghai Institute of Applied Physics, Chinese
Academy of Sciences, Shanghai 201800, China}
\author{C. L. Zhou}
\affiliation{Shanghai Institute of Applied Physics, Chinese
Academy of Sciences, Shanghai 201800, China} \affiliation{Graduate
School of the Chinese Academy of Sciences, Beijing 100080, China}
\author{X. Z. Cai}
\author{D. Q. Fang}
\author{W. D. Tian}
\author{H. W. Wang}
\affiliation{Shanghai Institute of Applied Physics, Chinese
Academy of Sciences, Shanghai 201800, China}

\date{\today}

\begin{abstract}
The detailed analysis of wide excitation function of nuclear stopping
has been studied within a transport model, Isospin-dependent Quantum
Molecular Dynamics model (IQMD) and an overall good agreement with the
INDRA and FOPI experimental data has been achieved. It is found that mean value of
isotropy in central Heavy-Ion Collision (HIC) reaches a minimum near
Fermi energy and approaches a maximum around 300 - 400$A$ MeV. This
suggests that, in statistical average, the equilibration is far from being
reached even in central HIC especially near Fermi energy. A
hierarchy in stopping of fragments, which favors heavy fragments
to penetrate, provides a robust restriction on the global trend of nuclear
stopping and could serve as a probe for nuclear equation of state.

\end{abstract}

\pacs{ 25.70.-z, 21.65.Mn}

 \maketitle
\section{Inroduction}

The equation of state (EOS)  of nuclear matter and its transport
mechanisms are the focus of Heavy-ion Collisions (HIC) during the
past three decades
\cite{Danil,Li2008,DiToro,Tsang,Nato,Hartnack2006,Sturm2001,Bord}.
Before the early 1980s, hydrodynamics approaches based on the
local equilibration postulate were provided to extract the EOS
information from experimental observables \cite{Sto}. The
discovery of collective flow seems to confirm these approaches and
their corresponding equilibration conditions. With the
development of microscopic transport theories, it is found that
the local equilibration postulate is not indispensable. Actually,
near Fermi energy, both statistical and dynamical models
which are based on very different, even contradictory mechanisms
can predict the same multifragmentation phenomenon
\cite{LeFvre2008}.

During central HIC process, the nuclear stopping governs most of
the dissipated energy and constrains the different reaction
mechanisms at different incident energies. It can provide the
information on the EOS, nucleon-nucleon (N-N) cross section and
the degree of the equilibrium reached in HIC
\cite{Reisdorf2010,Lehaut2010,Liu2001,Cao2010,Danielewicz2009,Gaitanos2004,Reisdorf2004,Zhang2007,Andronic2006,Kumar2010}.
Recently, INDRA and ALADIN Collaborations investigated
event-by-event nuclear stopping in central HIC at intermediate
energies by 4$\pi$ multidetector \cite{Lehaut2010}. In the
experiment \cite{Lehaut2010}, a striking minimal stopping at 40$A$
MeV (Fermi energy) for Xe+Sn system was observed. These
experimental results are potential to provide important
information about the properties of the nuclear matter and help to
find out critical clues for the whole dynamical process during
HIC. In previous comparison between experimental results and one
isospin-dependent Quantum Molecular Dynamics model (IQMD)
simulations \cite{Liu2001}, a significant difference between them
is observed. In the experiment \cite{Lehaut2010}, below 70$A$ MeV,
they show the experimental data far below the IQMD simulation
results and argued that IQMD model overpredicts nuclear stopping
power at low energy. However, before comparing the experimental
results quantitatively with that of transport model, two important
ingredients should be carefully considered. One is the difference
between nucleon phase space (the momentum and positions ensemble
of all nucleons) and fragment phase space (the momentum and
positions ensemble of all fragments) in IQMD, and the other is the
criterion of the centrality in the experiment. After considering
these two gredients, we find that the nuclear stopping could be
served as a potential probe for determining the nuclear EOS.

This article is organized as following: After introducing some
details on IQMD model in Sec.II, two gredients are reconsidered in
Sec.III: one is the difference between nucleon phase space and
fragment phase space and the other is the impact parameter mixing.
Then we compare the simulation results of IQMD with INDRA's
experiments near Fermi energy as well as FOPI's at higher energy
in Sec.IV. Summary and conclusions are given in Sec.V.

\section{Isospin Dependent Quantum Molecular Dynamics (IQMD) Model}

The Quantum Molecular Dynamics (QMD) \cite{Aichelin1991} model,
is a n-body theory, to predict the behavior of HIC at intermediate
energies on an event by event basis.
The IQMD \cite{Hartnack1989,Hartnack1998} model, as inheriting from
the QMD model, considers the isospin effects at various aspects:
different density distribution for neutron and proton, the asymmetry
potential term in mean field, experimental cross-section for necleon-necleon
($\sigma_{np} \approx 3\sigma_{pp}, \sigma_{pp}=\sigma_{nn}$) and Pauli-blocking
for neutron and proton separately. Just like in QMD, each nucleon
wave function is represented as a Gaussian form in IQMD, which defined as:
\begin{equation}
\label{eq:gauspack}
  \phi_i(\vec{r},t) = \frac{1}{{(2\pi L)}^{3/4}}
exp[-\frac{{(\vec{r}- \vec{r_i}(t))}^2}{4L}] exp[-\frac{i\vec{r}
\cdot \vec{p_i}(t)}{\hbar}].
\end{equation}
Here $L$ is the width parameter for the Gaussian wave-packet,
which is found to be dependent on the size of the reacting system,
to constrain the stability of system. We use $L$ = 2.16 ${\rm
fm}^2$ for Xe+Sn system. The $\vec{r_i}(t)$ and  $\vec{p_i}(t)$
are the position and momentum coordinates of nucleon. Performing
variation method, equations of the time evolution of the mean
position $\vec{r_i}(t)$ and momentum $\vec{p_i}(t)$ are found to
be well known Hamilton equations of motion:
\begin{equation}
\dot{\vec{p}}_i = - \frac{\partial \langle H \rangle}{\partial \vec{r}_i}
\quad {\rm and} \quad
\dot{\vec{r}}_i = \frac{\partial \langle H \rangle}{\partial \vec{p}_i} \,,
\end{equation}
where $\langle H \rangle$ is the Hamiltonian of the system.
So now the problem is to calculate the Hamiltonian of the system.

After applying the Wigner transformation to the single nucleon wave function,
one can get Wigner distribution fuction to describe a single nucleon density
in phase space defined as:
\begin{equation} \label{wignerfuction}
 f_i (\vec{r},\vec{p},t) = \frac{1}{\pi^3 \hbar^3 }
 {\rm e}^{-(\vec{r} - \vec{r}_{i} (t) )^2  \frac{1}{2L} }
 {\rm e}^{-(\vec{p} - \vec{p}_{i} (t) )^2  \frac{2L}{\hbar^2}  },
\end{equation}
and the total density of the system is the sum over all the nucleons.

Then we can get total Hamiltonian in IQMD model:
\begin{equation}
\label{eq:hamiltonian}
\langle H \rangle = \langle T \rangle + \langle V \rangle,
\end{equation}
where the meanfield part is
\begin{equation}
\label{meanfield}
 \langle V \rangle = \frac{1}{2} \sum_{i} \sum_{j \neq i}
 \int f_i(\vec{r},\vec{p},t) \,
V^{ij}  f_j(\vec{r}\,',\vec{p}\,',t)\,
d\vec{r}\, d\vec{r}\,'
d\vec{p}\, d\vec{p}\,'. \quad
\end{equation}
In QMD model, the two-body potential interaction contains the Coulomb interaction
and the real part of the G-Matrix. The later one can be divided
into three parts: the contact Skyrme-type interaction, a finite range
Yukawa-potential, and a momentum dependent part.
Mathematecally, the two-body potentail interaction can be written as:
\begin{eqnarray}
 \label{vijdef}
V^{ij} &=& G^{ij} + V^{ij}_{\rm Coul} \nonumber \\
       &=& V^{ij}_{\rm Skyrme} + V^{ij}_{\rm Yuk} + V^{ij}_{\rm mdi} +
           V^{ij}_{\rm Coul} \nonumber \\
       &=& t_1 \delta (\vec{x}_i - \vec{x}_j) +
           t_2 \delta (\vec{x}_i - \vec{x}_j) \rho^{\gamma-1}(\vec{x}_{i}) +\\
       & & t_3 \frac{\hbox{exp}\{-|\vec{x}_i-\vec{x}_j|/\mu\}}{
               |\vec{x}_i-\vec{x}_j|/\mu} +  \nonumber \\
       & & t_4\hbox{ln}^2 (1+t_5(\vec{p}_i-\vec{p}_j)^2)
               \delta (\vec{x}_i -\vec{x}_j) +
           \frac{Z_i Z_j e^2}{|\vec{x}_i-\vec{x}_j|}, \nonumber
\end{eqnarray}
where $Z$ is the charge of the baryon,  $t_1...t_5$ and $\mu$ are the
parameters to fit the real part of G-matrix and the properties of nuclei.
For IQMD model, one additional part, the symmetry potential $V^{ij}_{sym}$,
is added to take into account the isospin effects.
The total potential in IQMD is then written as:
\begin{equation}
V^{ij}= V^{ij}_{\rm Skyrme} + V^{ij}_{\rm Yuk} + V^{ij}_{\rm mdi} +
           V^{ij}_{\rm Coul} + V^{ij}_{sym},
\end{equation}
where the symmetry potential
\begin{equation}
V^{ij}_{sym}= t_6 \frac{1}{\rho_0}
 T_{3i} T_{3j} \delta(\vec{r}_i - \vec{r}_j) \quad
\end{equation}
with $t_6$ = 100 MeV for fitting the Bethe-Weizs\"acker mass
formula, $T_{3}$ is the isospin third projection and $\rho_0$ is
the nuclear saturation density ($0.16 {fm}^{-3}$).

We focus on the Skyrme potential, the momemtum dependent potential,
due to their special connection to nuclear EOS.

One can fulfill the calculation for Skyrme potential
and momemtum dependent in Eq. \ref{meanfield} by
introducing the interaction density,
\begin{equation} \label{rhoint}
\rho_{\rm int}^i(\vec{r_i}) = \frac{1}{(4\pi L)^{3/2}}
\sum_{j \neq i} {\rm e}^{\displaystyle
-(\vec{r_{i}}-\vec{r_{j}})^2/(4L) }.
\end{equation}

The momentum dependent potential, which may be optional in QMD and IQMD, is
parameterized with the experimental data, can be written as
\begin{equation}
\label{mdipar}
U_{mdi} =        \delta \cdot \mbox{ln}^2 \left( \epsilon \cdot
                \left( \Delta \vec{p} \right)^2 +1 \right) \cdot
                        \left(\frac{\rho_{int}}{\rho_0}\right),
\end{equation}
where $\delta$ = 1.57 MeV and $\epsilon$ = 500  ${(GeV/c)}^{-2}$
are taken from the measured energy dependence of the
proton-nucleus optical potential \cite{Aichelin1987} and
$\rho_{int}=\sum{\rho_{ \rm int}^i(\vec{r_i})}$.

The Skyrme potential is
\begin{equation}
\label{skyene}
U^{\rm Sky} = \alpha(\frac{\rho_{int}}{\rho_{0}}) +
\beta{(\frac{\rho_{int}}{\rho_{0}})}^{\gamma},
\end{equation}
where $\alpha$, $\beta$ and $\gamma$ are the Skyrme parameters,
which connect tightly with the EOS of bulk nuclear matter.
After fitting the minimum binding energy and the compressibility
at the saturation density, one can get the parameters.
The nuclear compressibility, which is the second derivative of the
energy at the minimum with respect to the density is expressed as:
\begin{equation}
\kappa = 9 \rho^2 \frac{\partial^2}{\partial \rho^2}
\left( \frac{E}{A}\right) .
\end{equation}
$\kappa$ = 200 MeV means soft EOS, while $\kappa$ = 380 MeV is
for hard EOS.
Please see Table \ref{eostab} the Skymre sets of parameters for
different EOS.

\begin{table}
\begin{tabular}{c|cccccccc|}

 &$\alpha$ (MeV)  &$\beta$ (MeV) & $\gamma$ & $\delta$ (MeV) &$\varepsilon \,
 \left(\frac{c^2}{\mbox{GeV}^2}\right)$ \\
\hline
 S  & -356 & 303 & 1.17 & ---  & ---    \\

 SM & -390 & 320 & 1.14 & 1.57 & 500  \\

 H  & -124 & 71  & 2.00 & ---  & ---    \\

 HM & -130 & 59  & 2.09 & 1.57 & 500  \\

\end{tabular}
\caption{\label{eostab} Parameter sets for the nuclear equation of
state used in the
QMD model. S and H refer to  the soft (compressibility $\kappa=$200 MeV) and
hard equations of state (compressibility $\kappa=$380 MeV), M refers to the
inclusion of momentum dependent interaction. The results are taken from \cite{Hartnack1998} }
\end{table}

The other terms can be also calculated by applying the convolution
of the interaction with the Wigner distribution function.
After all the potential terms are determined, one can solve the
equations of motion of the baryons. For the mesons, only Coulomb
force is considered.

For the collisions, IQMD uses the experimental cross-section which
tell the isospin effects and nuclear medium effect \cite{Hartnack1998}.
To consider the fermion property, the Pauli-blocking is also adopted
after the collisions.

IQMD treats the many body state explicitly, contains correlation
effects to all orders and deals with fragmentation and fluctuation
of HICs natually. To recognize fragments produced in HICs, a
simple coalescence rule is used with the criteria $\Delta$r = 3.5
fm and $\Delta$p = 300 MeV/c between two considered nucleons.
Thus, nucleons dominated in Fermi motion will be constrained in
the fragments.

\section{Difference between Nucleon Phase Space and
Fragment Phase Space and impact parameter mixing}

To describe nuclear stopping power, the ratio of transverse to
parallel quantities is used in the experiment \cite{Lehaut2010}.
One definition is the energy-based isotropy ratio $R_{E}$,
\begin{equation}
\label{eq:Re}
  R_{E} = \frac{\sum E_{ti}}{2\sum E_{li}},
\end{equation}
where $ E_{ti} (E_{li})$ is the c.m. transverse (parallel) kinetic
energy and the sum runs over all products event by event. One can
expect $R_{E}=1$ for isotropy or full stopping, $R_{E}<1$ for partial
transparency, and $R_{E}>1$ for super stopping.
Another definition is $R_{p}$,
\begin{equation}
\label{eq:Rp}
  R_{p} = \frac{2\sum |p_{ti}|}{\pi\sum |p_{li}|},
\end{equation}
where momenta are used instead of energies.
However, they are actually the different forms of physics
realization of classical Maxwell distribution assumption.

At low energy, two-body collision between
nucleons is suppressed by Pauli-blocking greatly and the mean
field governs the HIC process. At freeze-out stage, the motion of
a large portion of nucleons is limited in certain fragments, in
which Fermi motion dominates over the others. While, at higher
energy, two-body collision comes to play the dominant role and
most of nucleons suffer violent collisions and get excited to
become free. In this case, the fragment phase space tends to get
closer to the nucleon phase space. While, in previous IQMD
simulation by Liu et al. \cite{Liu2001}, the nuclear stopping was
investigated from nucleon phase space at intermediate energy, where
the isotropic Fermi motion still plays a great role. In comparison
with the experimental data of the
stopping, the fragment phase space should be applied at
intermediate energy within IQMD model rather than nucleon phase space.

\begin{figure}
\centering
\setlength{\belowcaptionskip}{-0.3cm}
\includegraphics[width=8.6cm]{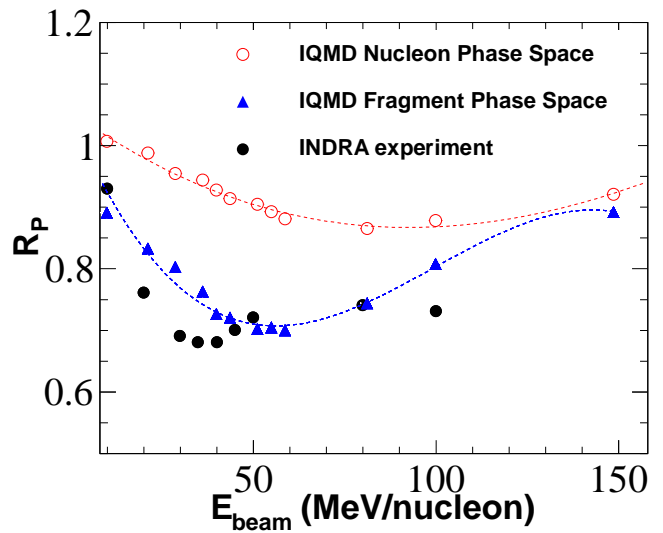}
\vspace{-0.1truein} \caption{\footnotesize (Color online)
Comparison between experimental stopping values \cite{Lehaut2010}
(black points) and predictions of IQMD calculation in nucleon
phase space (red open circle) and fragment phase space (blue up
triangle) for central $^{129}$Xe + $^{120}$Sn collsion. Two lines
are for guiding the eyes. IQMD adopts the hard EOS and
$b_{red}<0.1$. }\label{figRp_E}
\end{figure}

In Fig.~\ref{figRp_E}, we show the results of stopping defined as
the ratio of transverse momenta to parallel momenta $R_{p}$
, in nucleon phase space and fragment phase space, respectively.
The scaled impact parameter, $b_{red}<0.1$
($b_{red}=b/b_{max}$) is adopted, with $b_{max} = 1.12 (A_{P}^{1/3}
+ A_{T}^{1/3})$~fm, where $A_{P}$ and $A_{T}$ are the mass of projectile
and target, respectively.
Apparently, stopping calculated in nucleon phase space reaches almost
the saturated value, just like Liu's situation \cite{Liu2001}.
While, in the case of fragment phase space, stopping become far below
the saturated value and get closer to the experimental value.
It underlies that, at low incident energy, nuclear stopping obttained from
fragment phase space is more sensitive as compared to the nucleon phase
space and tends to fill the gap between theory and expeiments.
At higher incident energy (above 100$A$ MeV), the difference between stopping
presented in nucleon phase space and fragment space in IQMD, tends
to get smaller. However, a minimum stopping with value 0.70 at
about 55$A$ MeV is shown in IQMD, compared that with value 0.68 at 40$A$ MeV
in experiment. This difference may come from the different criteria
of centrality between IQMD and experiment.

To avoid autocorrelations, total charged particle multiplicity
($N_{ch}$) in the experiment \cite{Lehaut2010} is chosen as the
criterion of the centrality. However, the largest $N_{ch}$ does
not mean the highest centrality, due to event-by-event fluctuation.
On one hand, $N_{ch}$ gets
saturated at certain centrality and then central and near central HIC
provide similar $N_{ch}$. On the other hand, due to the dynamical
fluctuation, $N_{ch}$ at certain centrality presents a broad
distribution. Thus, the HIC at different centralities may share
the same $N_{ch}$ with a great chance. Hence, if $N_{ch}$ is adopted
for the criterion of the centrality, it is expected that the impact
parameters will suffer a great mixing and cover a wide range,
which will decrease the stopping power and increase the
fluctuation of the HIC system greatly.

\begin{figure}
\includegraphics[width=8.6cm]{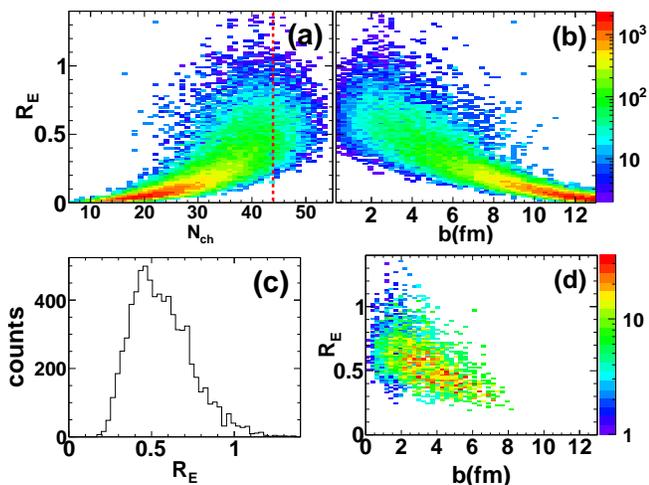}
\vspace{-0.35truein} \caption{\footnotesize (Color online) IQMD
 simulation with the hard EOS for $^{129}$Xe + $^{119}$Sn collisions at
50$A$ MeV. (a) Bi-dimensional correlation plot between the mean
value of stopping and the total charged particle multiplicity,
same as Fig.1(b) in Ref.~\cite{Lehaut2010}. (b) Bi-dimensional
correlation plot between the mean value of stopping  and impact
parameter. (c) Distribution of $R_{E}$ for central collisions. (d)
same as (b), but with central collision cut. }\label{figRe_Nch_b}
\end{figure}

We now use IQMD model to elucidate the impact parameter mixing
situation. In the experiment \cite{Lehaut2010}, a typical
cross-section of 50 mb for the most total charged particle
multiplicity  is taken for all HIC systems. Within the IQMD model,
we calculate the mean $N_{ch}$ and its width at head-on collision
($b_{red} = 0$). Then, low limit is set as the mean $N_{ch}$ with
its half width off. In Fig.~\ref{figRe_Nch_b}(a), IQMD model reproduces
well the experimental results of INDRA and ALADIN Collaborations
\cite{Lehaut2010}, except giving a little higher $N_{ch}$
(note that the experimental filter is not used in the
present calculation). The red dash line at $N_{ch}
= 44$ is the low limit for central collision in IQMD which is
comparable with $N_{ch} = 36$ in experiment. In
Fig.~\ref{figRe_Nch_b}(b), a mean value of stopping ($R_{E}$)
around 0.6-0.7 is observed in the exact central collision, which has
a decreasing trend with the increase of impact parameter.
When the cut $N_{ch} = 44$ is applied, there is a broad distribution
of impact parameter
in Fig.~\ref{figRe_Nch_b}(d), and a lower mean value of stopping
about 0.56 with a width about 0.42 in Fig.~\ref{figRe_Nch_b}(c),
which is comparable with that a mean value 0.56 with a width 0.47
in experiment \cite{Lehaut2010}. In this context,  better
criterion of centrality should be provided to reduce the
probability of impact parameter mixing at low energy.

\section{restuls and discussions}

Taking the above two key ingredients into account, we can
compare the experimental results with those of IQMD
quantitatively. To have a broad view of the nuclear stopping, a
systematic simulation of $^{129}$Xe + $^{120}$Sn is made in very
central HIC ($b_{red}<0.15$) from 10$A$ MeV to 1.2$A$ GeV to cover
most of the intermediate energy range, where our
results can be compared with the recent INDRA and FOPI data
\cite{Reisdorf2004,Reisdorf2010}.

It should be noted that related but different definitions of
stopping are adopted by INDRA \cite{Lehaut2010} and FOPI
\cite{Reisdorf2004,Andronic2006,Reisdorf2010} Collaborations,
respectively. The event-by-event stopping (event level) defined as
the ratio of transverse to parallel kinetic energies is adopted by
INDRA and ALADIN Collaborations (i.e. $R_{E}$). While, the ratio
of the variances of the transverse to that of longitudinal
rapidity distribution of fragment (fragment level) is taken in the
case of FOPI experiment
\begin{equation}
\label{vartl}
vartl =\frac{(\Delta y_{t})^{2}}{
(\Delta y_{l})^{2}}.
\end{equation}
Despite this, they are the same in spirit
to show the global stopping power of central HIC. When a full
stopping stage (likely to be in equilibration) is reached, both
$R_{E}$ and $vartl$ are required with a value of unit.

\begin{figure}
\includegraphics[width=8.6cm]{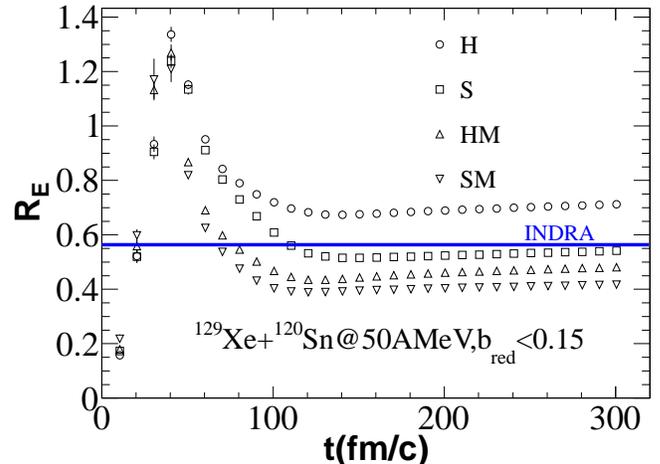}
\vspace{-0.3truein} \caption{\footnotesize (Color online)
Energy-based stopping values evolution with time for $^{129}$Xe +
$^{120}$Sn at 50$A$ MeV and $b_{red}<0.15$ in IQMD (open symbols)
and experimental data (solid line) \cite{Lehaut2010}. Circles,
squares, up-triangles and down-triangles represent our
calculations with the hard (H), the soft (S), the hard with MDI
(HM), and the soft with MDI (SM) for EOS, respectively. The
horizontal line represents the measured result by the INDRA
collaboration (not for the time evolution). }\label{Re_t}
\end{figure}

Now we compare our results with INDRA experiment. The central
($b_{red}<0.15$) collision of $^{129}$Xe+$^{120}$Sn, at incident
energy 50$A$ MeV is taken as an example. The time evolution of the
stopping (defined in $R_{E}$) for various EOS are shown in
Fig.~\ref{Re_t}. The stopping for different EOS presents the same
trend. The stopping rises up at the early compression stage, drops
down at the following expansion stage and becomes stable after the
system reaches the freeze-out. This is consistent with the results
from works of \cite{Liu2001} and \cite{Cao2010}, which are in
nucleon phase space. However, the stopping power shows EOS
dependence. The hard EOS shows higher stopping than the soft EOS,
during the whole time evolution. This is due to the hard EOS
nuclear matter is harder to be compressed than the soft one. When
compressed from the longitudinal direction, the hard EOS nuclear
matter will squeeze out more than the soft one, through the
tranverse expansion. The momentum dependent force decreases the
stopping power of the system, this may result from its repusive
nature and increasing in mean free path \cite{Aichelin1987}. At
this incident energy, the INDRA experimental result seems to
support the soft EOS.

It may be pointed out that here the time adopted in IQMD for the
freeze-out stage is 200fm/c. At low incident energy case, the
heavy fragments may still stay at an exciting state, which would
deexcite themselves by emissing free nucleons, LCP and gamma rays
at the later stage. Basing on this consideration, some hybrid
models, a dynamical one followed by a statistical one, come into
the play, such as AMD + GEMINI~\cite{Ma2002,Chen2010,Huang2010}
and HIPSE + SIMON ~\cite{Lacroix2004} etc. They do fit the
experimental fragment distribution very well. At the same time,
the deexcitation procedure may also extend both transverse and
longitudinal width of rapidity of the final fragments, which
smears the original stopping at fragment level. Especially, the
big fragments at high excitation stage will also emit free
nucleons or LCPs, which will increase the isotropy of the system.
This might be very serious for lower energy ($\leq$40$A$MeV) case
due to more heavy fragments are left. As a compensation, the
freeze-out time is delayed from 200fm/c to 300fm/c (or even
later), but no great changes happen, except for the IMF raise
their stopping value a little higher. For higher energy ($\geq$
100$A$ MeV), since there exists very few heavy fragments, it is
not necessary to evolute the system with such a long time or
deexcite the system with a statistical process ~\cite{Muller}.

\begin{figure}
\includegraphics[width=8.6cm]{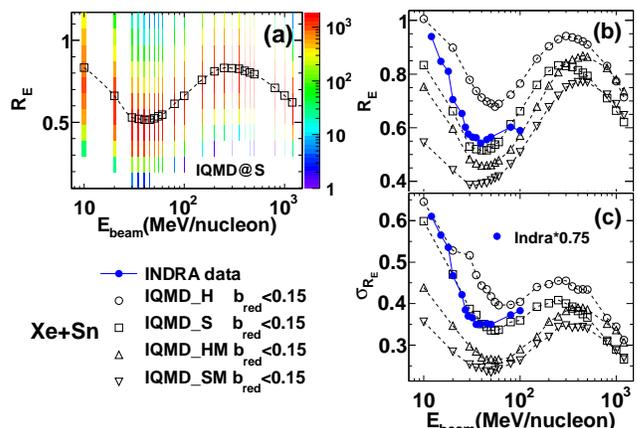}
\vspace{-0.3truein} \caption{\footnotesize (Color online) Stopping
and its width as a function of incident energy with different EOS
for central $^{129}$Xe + $^{120}$Sn collision in IQMD (open
symbols) and experimental data (solid circles) \cite{Lehaut2010}.
(a) Bi-dimensional plots show the distribution of $R_{E}$. (b) The
mean value of stopping. (c) The width of stopping  (Note the
widths of experiment have been scaled by a factor 0.75). Circles,
squares, up-triangles and down-triangles represent our
calculations with the hard (H), the soft (S), the hard with MDI
(HM), and the soft with MDI (SM) for EOS, respectively.
}\label{figRe_E}
\end{figure}

Fig.~\ref{figRe_E} shows the excitation function of the mean value
of  $R_{E}$ and its width $\sigma_{R_{E}}$ for
$^{129}$Xe+$^{120}$Sn by using different EOS. In
Fig.~\ref{figRe_E}(a), the soft EOS with the compressibility of
$\kappa$ = 200 MeV is taken to show the distribution of $R_{E}$
with incident energies. In Fig.~\ref{figRe_E} (b) and (c), the
mean $R_{E}$ and its width $\sigma_{R_{E}}$ show the same trends,
consistent with the experimental results \cite{Lehaut2010}. Scaled
by 0.75, the width of stopping in the experiment of INDRA is
compared with the simulation results. The broader widths in
experiment could come from the impact parameter mixing, as we have
discussed in Sec.II. For all the considered energy range, the hard
EOS with $\kappa$ = 380 MeV shows stronger stopping than the soft
EOS. The EOS with momentum dependent interaction (MDI)
\cite{Aichelin1987} tends to decrease the stopping at energy below
500$A$ MeV. While the weight of MDI tends to vanish above 500$A$
MeV. Near the Fermi energy, we also observe a minimum of stopping
with its minimal width for all EOS. The EOS with MDI favors more
penetration and consumes less energy to reach the minimal
stopping. The present experimental results seem to favor the soft
EOS around Fermi energy. This also consists with recent conclusion
on the soft hadronic matter basing on Kaon spectrum analysis at
higher energy \cite{Hartnack2006,Sturm2001}. At 300 - 400$A$ MeV,
a maximum of stopping accompanied by its maximal width is seen for
all EOS. This is comparable to the recent experimental results of
FOPI collaboration \cite{Reisdorf2004,Andronic2006,Reisdorf2010}.

\begin{figure}
\includegraphics[width=8.6cm]{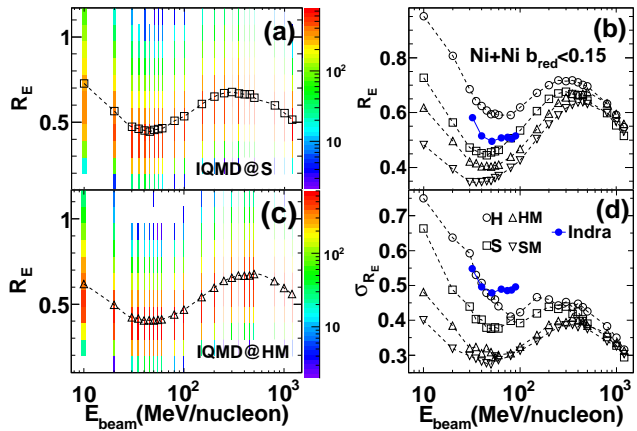}
\vspace{-0.3truein} \caption{\footnotesize (Color online) Same as
Fig.~\ref{figRe_E}, but for $^{58}$Ni + $^{58}$Ni system.
One more case (HM), is added in (c) to see the distribution of $R_{E}$.
}\label{figRe_E_Ni}
\end{figure}

The smaller size collision system $^{58}$Ni + $^{58}$Ni is also
investigated to see the stopping behavior and similar behaviors
are found in comparison with the case of $^{129}$Xe + $^{120}$Sn,
except for $^{58}$Ni + $^{58}$Ni shows lower stopping power. It
may be noticed again, the INDRA data seems to support the soft EOS
case. (Fig.~\ref{figRe_E_Ni}(b))

\begin{figure}
\includegraphics[width=8.6cm]{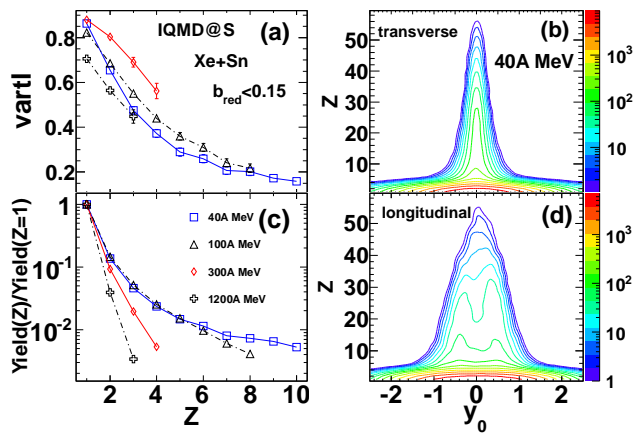}
\vspace{-0.3truein} \caption{\footnotesize (Color online) Stopping
hierarchy in IQMD with the soft EOS for central $^{129}$Xe +
$^{120}$Sn collision. Four energy  points, namely 40, 100, 300 and
1200$A$ MeV are investigated. (a) and (c) shows $Z$-dependent
stopping at fragment level and the yield of fragments (scaled by
Z=1), respectively. (b) and (d) displays bi-dimensional
distributions of transverse and longitudinal rapidity for
different fragments at 40$A$ MeV,
respectively.}\label{XeSn_b0.15_S_z_vartl}
\end{figure}

Fig.~\ref{XeSn_b0.15_S_z_vartl}(a) displays the atomic number
($Z$) hierarchy of the degree of stopping at fragment level at
different incident energies. From small fragment  $Z = 1$ with
stopping value $0.7-0.9$ to intermediate one $Z = 10$ with
$0.1-0.2$, indicates that the stopping power drops quickly as
$Z$ of the fragment increases. Similar trend has been experimentally
observed Au + Au, Xe + CsI and Ni + Ni central collisions at
150$A$ MeV, 250$A$ MeV and 400$A$ MeV, by FOPI collaboration \cite{Andronic2006,Zbiri2007,Reisdorf2010}.

The hierarchies also show the incident energy dependence. On the
one hand, at 300$A$ MeV (open diamond in
Fig.~\ref{XeSn_b0.15_S_z_vartl}(a)), fragments show stronger
stopping than those at the others which is consistent with the
maximum event-level stopping at the same energies in
Fig.~\ref{figRe_E}. This may suggest a state close to
equilibration. On the other hand, at 40$A$ MeV (open square in
Fig.~\ref{XeSn_b0.15_S_z_vartl}(a)) where the stopping at event
level reaches a minimum (Fig.~\ref{figRe_E}(b)), the stopping at
fragment level does show its corresponding smallest value. At this
incident energy, most of nucleons are confined in certain
fragments, which also reduces the stopping value at event level.
Near Fermi energy, the relative yields of intermediate-mass
fragments (scaled by $Z=1$) are higher than those at the others as
shown in Fig.~\ref{XeSn_b0.15_S_z_vartl}(c), which will lead to
the smallest effective power-law exponents in charge distribution
~\cite{Ma2005}. For a further understanding the minimal stopping
near Fermi energy, in Fig.~\ref{XeSn_b0.15_S_z_vartl}(b) and
Fig.~\ref{XeSn_b0.15_S_z_vartl}(d), the holographic 2D histogram
of transverse and longitudinal rapidity for different charge of
fragments are presented (rapidity has been scaled by the
projectile rapidity in center of mass system). From $Z=1$ to
larger $Z$, the longitudinal rapidity distributions become
relative broader than the transverse one. In addition, for those
fragments with $Z$ larger than about 4, two clear peaks in the
longitudinal rapidity stand prominently, corresponding to
projectile-like and target-like contribution. This means at Fermi
energy, equilibration is far from being reached and entrance
channel effects remain strong even in central HIC. While, at high
energy 1.2$A$ GeV, fragments are the most transparent.

\begin{figure}
\includegraphics[width=8.6cm]{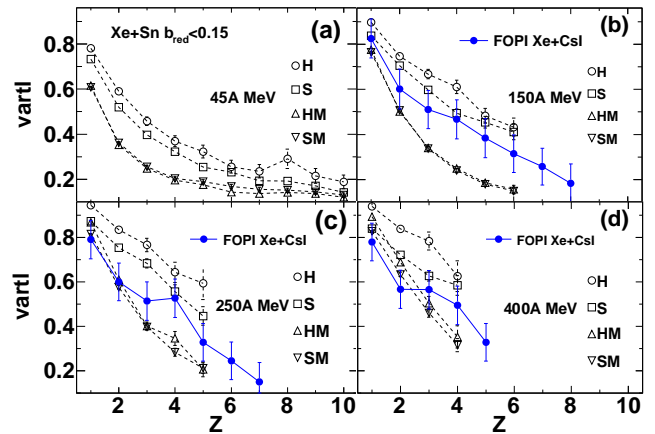}
\vspace{-0.2truein} \caption{\footnotesize (Color online) Atomic
number dependence of the stopping observable $vartl$ for fragments
emitted in central  $^{129}$Xe + $^{120}$Sn collisions. Open
symbols represent our calculations with different EOS and solid
symbols with error bar denote the available experimental data
\cite{Reisdorf2010}. (a) 45$A$ MeV (no experimental data available
yet); (b)  150$A$ MeV; (c)  250$A$ MeV; (d) 400$A$ MeV.
}\label{hierarchy_eos}
\end{figure}

It is highly expected that the hierarchy would help to determine
nuclear EOS. Fig.~\ref{hierarchy_eos} shows $Z$-dependent stopping
for four different energies with different EOS in IQMD. The hard EOS case
shows higher stopping at fragment level than the soft EOS one, and the MDI
always makes it easy for the fragments to penetrate. At lower
energy (150$A$ MeV in Fig.~\ref{hierarchy_eos}(b)), the EOS
without MDI agrees with FOPI data \cite{Reisdorf2010} much better
(Xe+CsI are similar system as Xe+Sn). On the contrary, at higher
energy (Fig.~\ref{hierarchy_eos}(c)  and (d)), to reproduce
experimental data, the weight of MDI should be increased. More
information could be extracted once more experimental data are
provided to low down the statistical fluctuation. In addition, at
energy near Fermi energy as shown in Fig.~\ref{hierarchy_eos}(a),
it does deserve performing experiment to obtain  $Z$-dependent
stopping, which is important to understand the minimal stopping at
event level at Fermi energy as well as constrain the EOS.

The atomic number dependence of stopping observable suggests that
the heavier fragments have less stopping. This indicates that the
heavier fragments are  more transparent and keep a stronger entrance
channel memory even in the central nuclear collision. This
supports the argument that the fragments are not formed in a
globally equilibrated environment \cite{Aichelin1991}, which
contradicts the assumption of equilibrium, required by ideal fluid
hydrodynamics. A self-consistent explanation is provided in
\cite{Andronic2006}: if a heavier fragment can survive during the
process of HIC, its constituent nucleons should have suffered a
less violent collision history. Basing on this consideration, the
stopping hierarchy at fragment level would also reflect the N-N
cross section. This explanation are also quantitatively
demonstrated in earlier version of QMD \cite{Gossiaux1997}.
Another explanation is due to corona effect (geometrical) \cite{Reisdorf1997},
even in very central collsion, those nucleons near the surface of projectile
or target would suffer less collision history than those inside. Thus,
they would keep their original direction and compose the final
heavy residues.

In very central HIC,  the incident energy dependence of stopping
can be understood in a unified structure based on Pauli-blocking,
mean field and N-N collision. At intermediate energy, two-body
collision between nucleons always increases the dissipation of the
HIC system and tends to favor more isotropy. While, Pauli-blocking
suppresses N-N collision at low energy, and mean field forms
collective motion. This picture shows incident energy dependence.
Below Fermi energy, the HIC system has enough time to reach an
equilibration state because of the slow process of the reaction.
Near Fermi energy, N-N collision is suppressed by Pauli-blocking
greatly and nucleons tend to keep their original moving state and
have an entrance effect. Meanwhile, mean field dominates HIC, and
clusters are then favored. Above Fermi energy, two-body collision
between nucleons comes to play the dominant role and reach their
maximal effect at 300 - 400$A$ MeV. While, at higher energy, the
mean free path of nucleon increases because of the smaller N-N
cross section, thus most of nucleons will pass through with each
other.

\section{Summary and conclusions}
In summary, we successfully describe, for the first time,  the
wide-range excitation function of nuclear stopping from 10$A$ MeV
to 1.2$A$ GeV with a transport model IQMD. A minimum of nuclear
stopping value near Fermi energy and a maximum at about 300$A$ MeV
in very central HIC match the INDRA and FOPI data very well. The
former indicates that in statistical average, equilibration state
is far from being reached near Fermi energy even in very central
HIC. Meanwhile, the hierarchy of stopping observable together with
the yields of fragments, provides us a decomposition way to
understand the whole stopping excitation function.  Around  Fermi
energy, the soft EOS seems the best, while at energy from 250 to
400$A$ MeV, the role of MDI becomes  important. In addition, it is
also highly expected that the isospin which has been regarded to
reach its equilibration the fastest at the early HIC process might
also get a great penetrating in very central HIC near Fermi
energy.

\section{ACKNOWLEDGMENTS}
This work is partially supported by the NSFC under contract No.s
11035009, 10979074, 10875160, 10805067 and 10975174, the
973-Program  under contract No. 2007CB815004, the Shanghai
Development Foundation for Science and Technology under contract
No. 09JC1416800, the Knowledge Innovation Project of Chinese
Academy of Sciences under Grant No. KJCX2-EW-N01.

\end{document}